\documentclass[journal]{IEEEtran}

\usepackage{amsmath,amssymb,amsthm}
\usepackage{cite}
\usepackage{color}
\usepackage{graphicx}
\usepackage{hyperref}
\usepackage{multirow}
\usepackage[normalem]{ulem}
\usepackage{url}

\begin{document}

\pagenumbering{gobble}

\title{Delay Aware Intelligent Transient Stability Assessment System}

\author{James J.Q. Yu, \IEEEmembership{Member,~IEEE},
		Albert Y.S. Lam, \IEEEmembership{Senior Member,~IEEE},
		David J. Hill, \IEEEmembership{Life Fellow,~IEEE},
        and Victor O.K. Li, \IEEEmembership{Fellow,~IEEE}
\thanks{The authors are with the Department of Electrical and Electronic Engineering, The University of Hong Kong, Pokfulam Road, Hong Kong (e-mail: \{jqyu,ayslam,dhill,vli\}@eee.hku.hk).

This work was supported by the Theme-based Research Scheme of the Research Grants Council of Hong Kong, under Grant No. T23-701/14-N.}
}

\markboth{IEEE Access}{\MakeLowercase{\textit{Yu}}: Delay Aware Intelligent Transient Stability Assessment System}

\maketitle

\begin{abstract}
Transient stability assessment is a critical tool for power system design and operation.
With the emerging advanced synchrophasor measurement techniques, machine learning methods are playing an increasingly important role in power system stability assessment.
However, most existing research makes a strong assumption that the measurement data transmission delay is negligible.
In this paper, we focus on investigating the influence of communication delay on synchrophasor-based transient stability assessment.
In particular, we develop a delay aware intelligent system to address this issue.
By utilizing an ensemble of multiple long short-term memory networks, the proposed system can make early assessments to achieve a much shorter response time by utilizing incomplete system variable measurements.
Compared with existing work, our system is able to make accurate assessments with a significantly improved efficiency.
We perform numerous case studies to demonstrate the superiority of the proposed intelligent system, in which accurate assessments can be developed with time one third less than state-of-the-art methodologies.
Moreover, the simulations indicate that noise in the measurements has trivial impact on the assessment performance, demonstrating the robustness of the proposed system.
\end{abstract}

\begin{IEEEkeywords}
Transient stability assessment, communication delay, long short-term memory, phasor measurement units, voltage phasor, intelligent system.
\end{IEEEkeywords}

\section{Introduction}

Transient stability refers to the capability of a power system to maintain its synchronism subject to large disturbances \cite{kundur_definition_2004}.
The transient stability issues caused by large disturbances are considered more serious than before as the power systems are being operated close to their stability limits to satisfy the increasing power demand \cite{demetriou_dynamic_????}.
Consequently, critical contingencies may lead to significant system failures or power blackouts.
In order to prevent such situations, system operators need to assess the stability condition of the grid and, when necessary, plan a collection of remedial control actions to retain the stability.
Therefore, transient stability assessment (TSA) in real-time is regaining interest from the community \cite{cepeda_real-time_2014}.

Many previous studies on TSA were conducted by using offline dynamic simulations for a collection of credible contingencies \cite{diao_design_2010}.
This methodology is widely adopted in designing the protective and control systems for secure operations.
Meanwhile, online TSA techniques are employed to evaluate the progress of transient dynamics of a power system in real time.

With the gradual adoption of synchrophasor measurement facilities, e.g., phasor measurement units (PMUs), a significant amount of effort has been devoted to utilizing real-time system variables for TSA decision making \cite{diao_design_2010,sun_online_2007,gao_decision_2011}.
On top of that, post-contingency remedial actions can be taken in real-time to give guaranteed TSA results \cite{cepeda_real-time_2014}.
With post-contingency system dynamics, techniques such as the piecewise constant current load equivalent method \cite{liu_application_1995} and emergency single machine equivalent method \cite{pavella_transient_2000} were proposed for online dynamic power system security assessment.
Machine learning techniques for TSA, on the other hand, received a lot of attention in recent years due to their relatively low assessment computational complexity.
Approaches like pattern recognition \cite{zhu_time_2016}, decision tree \cite{guo_probabilistic_2014}, artificial neural networks (ANN) \cite{hashiesh_intelligent_2012}, support vector machine \cite{guo_online_2016}, and fuzzy knowledge-based systems \cite{kamwa_accuracy_2012} were employed to realize fast-response online TSA.
For instance, our previous work \cite{yu_intelligent_????} handles TSA with a modern variant of ANN, and assessment results can be generated in an online manner.
These techniques extract the relationship between system variable measurements and their respective stability indices.
With this relationship, new transient stability dynamics can be assessed with minimal computational efforts.

It appears that all existing research on synchrophasor-based online TSA implicitly assumes that the wide area monitoring systems can provide reliable, accurate, and synchronized system variable measurements and there is no data transmission delay between PMUs and the central controller, see \cite{zhang_post-disturbance_2015, yu_intelligent_????} for examples.
However, although being measured in a synchronous manner thanks to PMUs, synchrophasors cannot reach the central controller in perfect synchrony due to unpredictable communication link congestion and routing delay \cite{zhang_service_1995}.
Therefore, it is necessary to design a data transmission delay-aware TSA system that can provide robust and reliable assessment results given delayed or missing system variable measurements \cite{he_online_2013}.

One feasible approach to address the delayed synchrophasor problem is to recover those missing data by learning the system model and other measurements when the delayed data packets are being tramsmitted \cite{gao_missing_2016,bai_measurement-based_2015}.
However, this approach suffers from a major drawback that the high computational complexity of these nonlinear state estimators results in a large delay for estimating the delayed synchrophasor, far from being commensurate with the PMU sampling rate \cite{he_online_2013}.
To overcome this drawback in the state estimation approaches, a data-mining based neural network ensemble prediction technique is utilized as an alternative in this paper.

This work focuses on establishing an intelligent system to address the transient stability assessment problem with time-delayed synchrophasors.
We construct the system using multiple advanced machine learning techniques and heuristics, namely Long Short-term Memory (LSTM) \cite{hochreiter_long_1997}, ensemble learning \cite{hansen_neural_1990}, rule-based decision machine heuristic, and two optimizers (Adam \cite{kingma_adam:_2015} for neural network training, and Social Spider Algorithm \cite{yu_social_2015} for non-convex optimizations).
Utilizing the advantages of involved techniques, the proposed system is able to make preliminary assessments at the earliest possible time and revise the predictions when more information is available.
Compared with previous TSA systems, the proposed one can achieve a faster response time to make accurate assessments.
In addition, the combination of multiple techniques contributes to the superior assessment performance as will be illustrated.

The remainder of this paper is organized as follows.
Section II introduces the delay aware TSA problem and its difference from tradition TSA.
Section III elaborates on the formulation and implementation of the proposed intelligent system.
Section IV demonstrates numerical results on a modified New England 10-machine test system.
Finally we conclude in Section V with a discussion of potential future research.

\section{Delay Aware Transient Stability Assessment} \label{sec:delay_TSA}

Typical stability assessment methodologies introduce the idea of observation window to facilitate their data collection process (See \cite[Section~2.1]{zhang_post-disturbance_2015} for a detailed introduction).
After the clearance of a fault, the power system dynamic behavior is observed for a certain period of time by PMUs, and this information is later utilized to make a stability prediction for the future.
While some recent work has improved the TSA system by replacing the block observation with continuous observation \cite{zhang_post-disturbance_2015}, it still relies on an implicit but strong assumption: the central controller needs to receive measurements with the same timestamp from different data sources simultaneously.

From the communication perspective, however, it is not realistic to postulate that no data transmission delay is incurred over such communication links \cite{zhang_service_1995}.
Suppose the system measurements from PMU $p$ reach the central controller at time $t+\Delta t_{p,t}$, where $t$ is the time index after the fault clearance, and $\Delta t_{p,t}$ is the data transmission delay induced by the message communication from $p$ at $t$.
The previous TSA work generally assumes $\Delta t_{p,t}\equiv0$.
However, investigations on data transmission protocols demonstrate that modeling such a delay is a complicated task and cannot be simply replaced with a constant \cite{gupta_capacity_2000}.
Moreover, the measurements actually follow an asynchronous and disordered pattern: $\Delta t_{p,t}<\Delta t_{q,i}$ has no direct implication on $\Delta t_{p,t+1}<\Delta t_{q,t+1}$, where $p$ and $q$ are arbitrary PMUs.

According to the IEEE Standard for Synchrophasor Data Transfer for Power Systems (C37.118.2-2011) \cite{_ieee_2011}, typical values for the transmission delay between PMU and Phasor Data Concentrator (PDC) are between 20 ms to 50 ms, and this delay may be further increased due to temporary data congestion over the communication links.
The delay should be added to the system response time the TSA calculation time to give, which obstructs the early adoption of subsequent control actions.
Therefore, research needs to be carried out on proposing mechanisms that can make transient assessments without waiting for the arrival of all PMU measurements.
Such mechanisms should develop a similar TSA accuracy compared with the state-of-the-art TSA algorithms, while making assessments far earlier than those made with full communication delays.
In this paper we aim to devise an intelligent delay aware TSA system to achieve this objective.
The proposed system copes with the asynchronous arrival of synchrophasor data, and minimizes their influence on the system response time.

\section{Proposed Delay Aware Assessment System}

In this section, we elaborate on how we design an intelligent system to perform TSA considering delayed synchrophasor data packets arriving in an asynchronous manner.
We will first introduce Long-short Term Memory (LSTM) networks, and the reason why we use such networks to construct the system.
Then we overview the structure of the system followed by the illustrations of other building blocks and the decision machine.
Finally we will explain the overall work flow of the proposed delay aware TSA system.

\subsection{An Introduction to Long Short-term Memory}\label{sub:LSTM}

ANN is one of the automatic machine learning techniques and it has been employed in a variety of disciplines in the last few decades \cite{lippmann_introduction_1987}.
One major merit of this technique that makes it suitable for online data processing is its excellent responsiveness \cite{lippmann_introduction_1987}.
ANN functions like a black box.
By training with supervised datasets, ANN learns the mathematical relationship of inputs $X$ and outputs $Y$, or their distributions \cite{narendra_identification_1990}.

LSTM is a variant of ANN \cite{hochreiter_long_1997}.
From the functional point of view, LSTM differs from the conventional ANN in that it considers the temporal data correlation in $X$.
In addition, LSTM generally outperforms other time-dependent variants of ANN, e.g., recurrent neural network, as it avoids the occurrence of the ``vanishing gradient problem'', which deteriorates the temporal correlation extraction efficiency \cite{hochreiter_long_1997}.

\begin{figure}
\centering
\includegraphics[width=\linewidth]{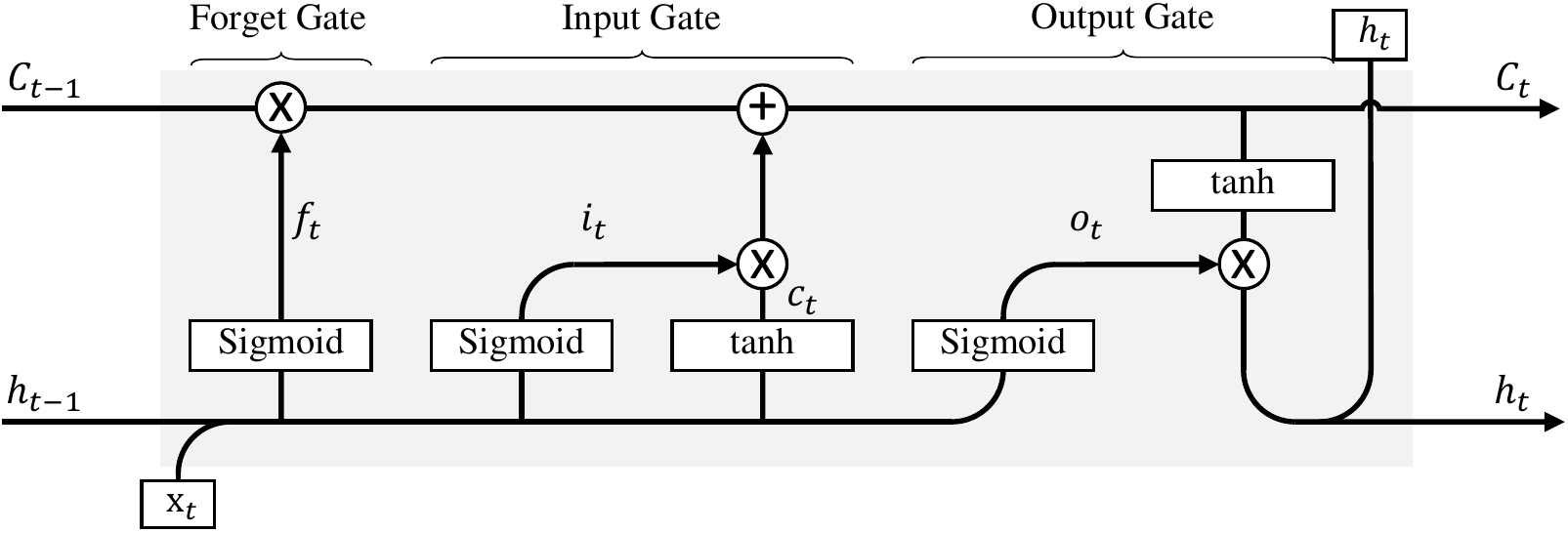}
\caption{The LSTM memory block with the memory cell $C_t$.}
\label{fig:LSTM}
\end{figure}

\begin{figure}
\centering
\includegraphics[width=\linewidth]{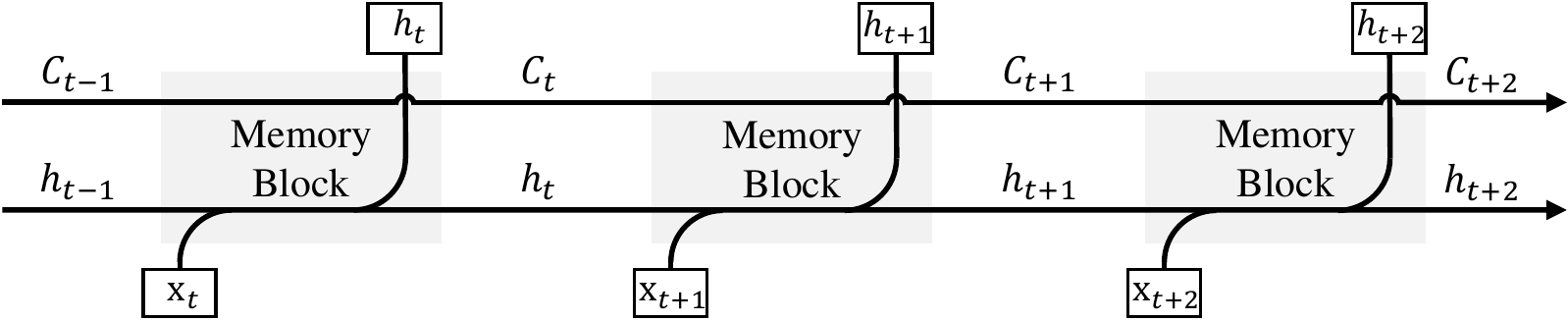}
\caption{Unrolled form of a typical LSTM network.}
\label{fig:LSTM_unroll}
\end{figure}

Fig. \ref{fig:LSTM} illustrates one memory block of a typical LSTM network.
This block utilizes the input data with a specific timestamp as well as the memory from the previous timestamp for feature extraction.
The processed information, or memory, is stored in the memory cell $C_t$ and passed in the next time slot \cite{hochreiter_long_1997}.
The output data $h_t$ is also generated based on $C_t$.
This process is depicted in Fig. \ref{fig:LSTM_unroll}.

As illustrated, the LSTM memory block comprises three gates, namely, the forget, input, and output gates; the forget and input gates manage the existing network memory and the new input information, while the output gate controls the output information.
To compute $C_t$, these gates manipulate the temporal data correlation stored in the LSTM network as follows:
\begin{IEEEeqnarray}{rCl} \IEEEyesnumber \label{eqn:input_forget_gate}
\mathbf{f}_t &=& \mathrm{Sigmoid}(\mathbf{W}_f\mathbf{x}_t+\mathbf{U}_f\mathbf{h}_{t-1}+\mathbf{b}_f) \IEEEyessubnumber\\
\mathbf{i}_t &=& \mathrm{Sigmoid}(\mathbf{W}_i\mathbf{x}_t+\mathbf{U}_i\mathbf{h}_{t-1}+\mathbf{b}_i) \IEEEyessubnumber\\
\mathbf{c}_t &=& \tanh(\mathbf{W}_C\mathbf{x}_t+\mathbf{U}_C\mathbf{h}_{t-1}+\mathbf{b}_C) \IEEEyessubnumber\\
\mathbf{C}_t &=& \mathbf{f}_t\ast \mathbf{C}_{t-1} + \mathbf{i}_t\ast \mathbf{c}_t. \IEEEyessubnumber
\end{IEEEeqnarray}
Consequently, the output is generated at the output gate, given by
\begin{IEEEeqnarray}{rCl} \IEEEyesnumber \label{eqn:output_gate}
\mathbf{o}_t &=& \sigma(\mathbf{W}_o\mathbf{x}_t+\mathbf{U}_o\mathbf{h}_{t-1}+\mathbf{b}_o) \IEEEyessubnumber\\
\mathbf{h}_t &=& \mathbf{o}_t\ast \tanh(\mathbf{C}_t). \IEEEyessubnumber
\end{IEEEeqnarray}
In (\ref{eqn:input_forget_gate}) and (\ref{eqn:output_gate}), $\ast$ is the element-wise product, $\mathrm{Sigmoid}(x)=(1+e^{-x})^{-1}$ is the sigmoid function, and $\mathbf{W}$, $\mathbf{U}$, $\mathbf{b}$ are matrices corresponding to the LSTM learning parameters.

With the LSTM memory blocks, neural networks can accept a sequence of inputs $\mathbf{x}_1, \mathbf{x}_2, \cdots, \mathbf{x}_t, \cdots, \mathbf{x}_T$ and develop timestamped outputs $\mathbf{h}_1, \mathbf{h}_2, \cdots, \mathbf{h}_t, \cdots, \mathbf{h}_T$, where $T$ is the length of input data.
Output $\mathbf{h}_t$ are generated with the complete existing knowledge from time 1 to $t$.
For instance, $\mathbf{h}_1$ is calculated solely using $\mathbf{x}_1$, and $\mathbf{h}_3$ is generated by $\mathbf{x}_1$, $\mathbf{x}_2$, and $\mathbf{x}_3$.
Utilizing this characteristic, LSTM can develop preliminary results once the first set of input $\mathbf{x}_1$ is available.
Therefore, the proposed LSTM-based system has the capability to generate very early TSA results, which can be later revised when more data is available.
It meets the delay aware TSA system requirements as stated in Section \ref{sec:delay_TSA}.

\subsection{Structure of Proposed Ensemble-based Intelligent System}

Ensemble of neural networks is a learning paradigm in which multiple neural networks are employed to solve a problem.
In statistical learning, it is widely recognized that such ensembles demonstrate improved generalization capabilities compared to standalone networks \cite{hansen_neural_1990}.
In an ensemble, classification errors of one single network can be compensated by others, to provide an enhanced robustness of the complete system.

\begin{figure}
\centering
\includegraphics[width=0.8\linewidth]{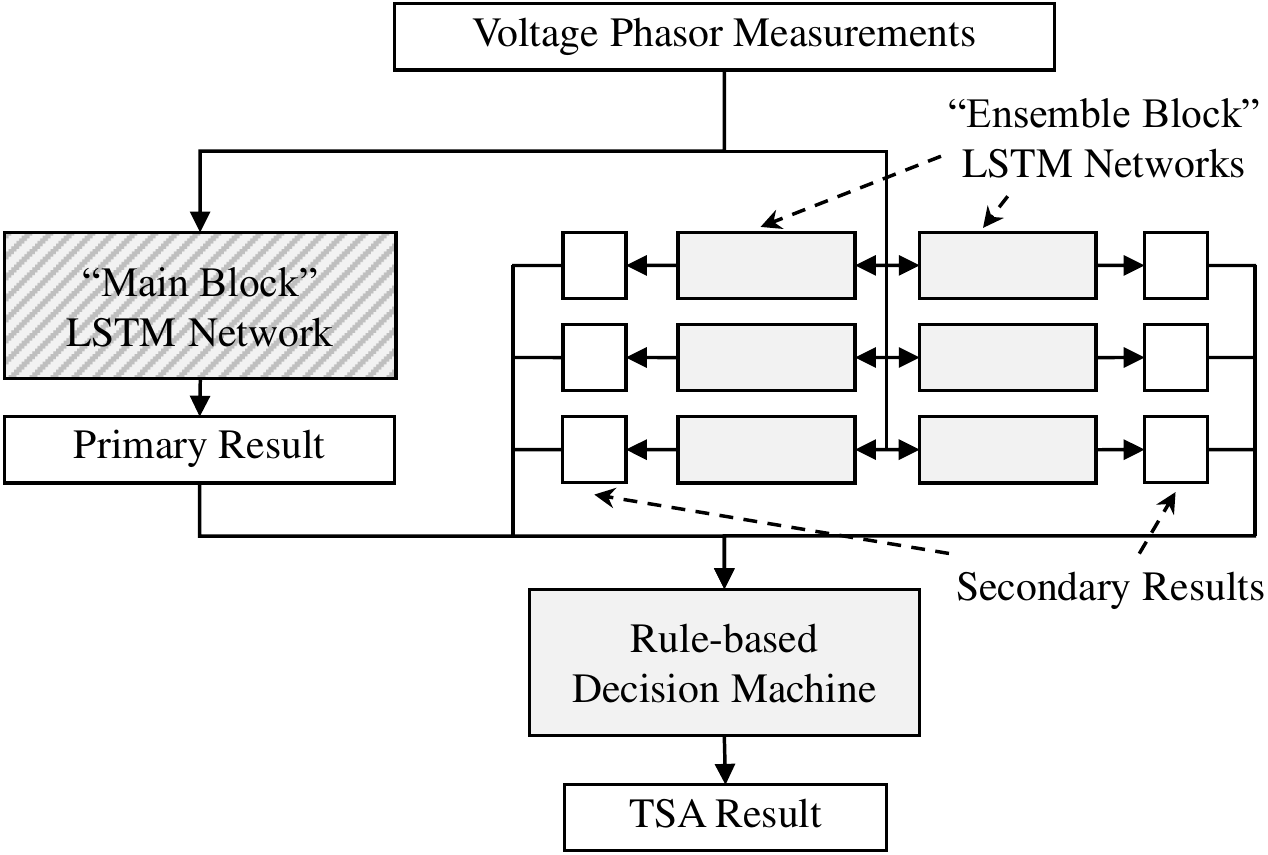}
\caption{Structure of the proposed ensemble-based intelligent system for TSA.}
\label{fig:IS}
\end{figure}

Due to the outstanding performance of neural network ensembles for classification tasks, we develop a novel LSTM ensemble-based intelligent system for efficient TSA.
The structure of such a system is shown in Fig. \ref{fig:IS}.
In the system, one \textit{Main Block} LSTM network (the striped block in Fig. \ref{fig:IS}) is employed to develop a \textit{Primary Result} using post-contingency bus voltage magnitude and angle measurements of the power system.
Meanwhile, $N$ \textit{Ensemble Block} networks generate $N$ \textit{Secondary results} in parallel.
While primary and secondary results share the same characteristics and purposes, we consider the primary result more reliable, as the main block has a larger network structure, which will be elaborated on in the following sub-sections, than individual secondary results.
These $N+1$ results are jointly considered in a rule-based \textit{Decision Machine} to produce the final TSA result.
The purpose of introducing this machine is to gather enough system stability assessment information from multiple sources before making a final conclusion.

Besides the different sizes of the main and ensemble blocks, their ways of addressing communication delays are also different.
The main block can develop TSA results despite the incompleteness of input data due to latency, but the primary results may sometimes be inaccurate.
The ensemble blocks will only develop their results when all required input measurements are known, and the secondary results are utilized to correct the primary result.
Consequently, when most measurements are not received, the main block can already gives preliminary assessments.
With the increase of received data, ensemble blocks start to correct the primary result if necessary.

\subsection{Main Block}

\begin{figure}
\centering
\includegraphics[width=0.65\linewidth]{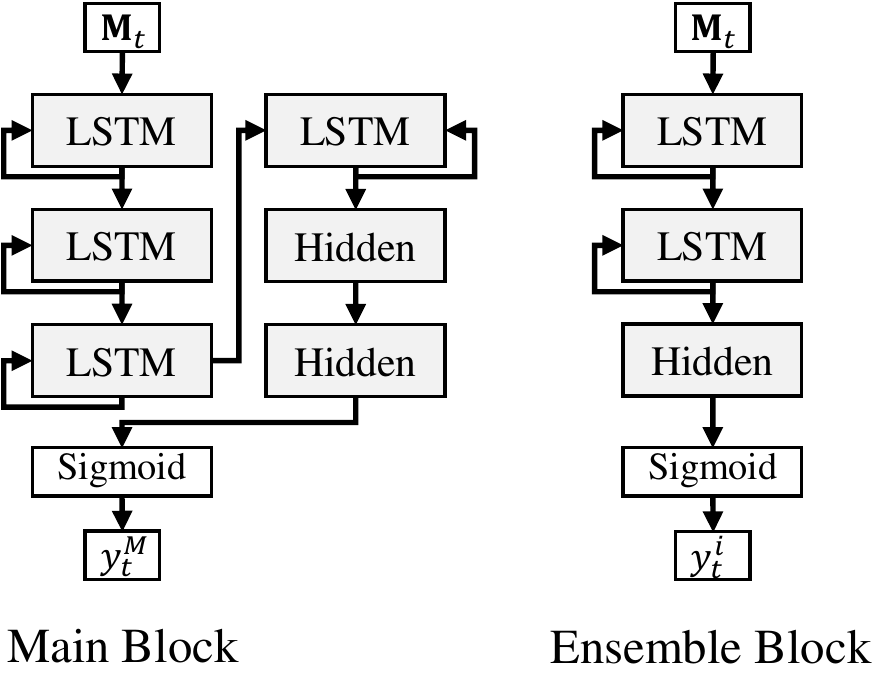}
\caption{Structure of the main block and ensemble blocks.}
\label{fig:Blocks}
\end{figure}

As illustrated in Fig. \ref{fig:IS}, the main block is employed to develop primary results for generating the final TSA result in the proposed system.
As this result is considered essential - this part will be further elaborated in Section \ref{sub:rule_machine} - we pay more attention to model its input-output dependency by employing a Deep Neural Network (DNN) with four layers of LSTM blocks and two fully-connected hidden neuron layers.
The structure of this DNN is depicted in Fig. \ref{fig:Blocks}.
While the LSTM blocks are capable of extracting the temporal data dependency from the input data, the neuron layers translate the extracted features to human-readable assessment results.
A final $\mathrm{Sigmoid}$ function is utilized to cast the results into $(0,1)$.
Owing to its superior capability of modeling highly non-linear relationships, this main block network is expected to output a more reliable result than the secondary results produced by the ensemble blocks, each of which contains significantly fewer number of layers.

\subsubsection{Offline Training Process}

Training the main block is conducted offline with all bus voltage phasor data obtained by contingency simulations.
Using 50/60-Hz sampling, the normalized measured voltage phasors are presented in the form of $\mathbf{M}^{B\times T}$, where $B$ is the number of buses.\footnote{In practice, it can be more economically efficient to install PMUs on a subset of buses in the power grid. In such cases, $\mathbf{M}^{B\times T}$ may also include ``measurements'' which are actually computed using power flow model.}
The input data $\mathbf{M}$ can be sliced along the time axis, resulting in vectors $\mathbf{M}_t=[M_{1,t}, M_{2,t}, \cdots, M_{B,t}]^\prime$, where $M_{b,t}$ is the system variable measurements of bus $b$ at timestamp $t$ after the fault clearance.
Each $\mathbf{M}_t$ comprises measurements from different buses with an identical timestamp.
The output stability assessment $y$ is obtained by observing the generation angle derivation and presented in 0-1 binary form.\footnote{In this paper, assessment result 1 means that the system will remain stable and 0 is unstable.}

Given a collection of $C$ training cases $\{\mathbf{M}_{(c)},y_{(c)}\}^C_{c=1}$ where $y_{(c)}$ is the assessment $y$ of case $c$, the training process aims to obtain the parameters of the LSTM \cite{hochreiter_long_1997} and fully-connected neuron layers \cite{lippmann_introduction_1987}.
In this paper, we employ the Adam optimizer \cite{kingma_adam:_2015} to find the optimal values for the above-mentioned parameters, and the binary cross entropy error function is selected as the objective function:
\begin{equation}
\text{minimize} -\sum_{c=1}^C[y_{(c)}\log y^M_{(c)}+(1-y_{(c)})\log(1-y^M_{(c)})],
\end{equation}
where $y^M_{(c)}$ is the actual assessment result of $\mathbf{M}_{(c)}$ with the main block.

\subsubsection{Online Assessment Process}

Online assessment cannot be conducted in the same way as training.
As introduced in Section \ref{sub:LSTM}, the proposed main block accepts one or multiple $\mathbf{M}_t$ vectors instead of an integral $\mathbf{M}$ block as input and outputs $y_t$.
However, some values in $\mathbf{M}_t$ can be unknown when the first several measurements in $\mathbf{M}_t$ reach the central controller.
In all previous work, e.g., \cite{cepeda_real-time_2014,diao_design_2010,sun_online_2007,gao_decision_2011,liu_application_1995,zhang_post-disturbance_2015}, the system may hang on until all synchrophasors in $\mathbf{M}_t$ become ready, but the waiting time ($\max\{\Delta t_{p,t}\}$) is significantly increased compared with premature assessments cases.

In this work, we employ a ``zero-padding'' scheme to pad all unknown values in $\mathbf{M}_t$ with zeros.
In addition, a user-defined parameter $\phi$ is introduced to determine whether a specific $\mathbf{M}_t$ is included in the input.
It is included only when 1) the percentage of known values in $\mathbf{M}_t$ is greater than $\phi$, and 2) $\mathbf{M}_{t-1}$ is also included when $t>1$.
In this design, the value zero in the input is considered ``unknown'', and the main block can start developing primary results much earlier than waiting for all synchrophasors.

During the online assessment process, one stability assessment result is developed whenever a new synchrophasor reaches the central control and $\mathbf{M}\ne\varnothing$.
This primary result is output to the decision machine to develop the TSA result.

\subsection{Ensemble Block}

While the main block is expected to develop considerably reliable assessment results, it is still inevitable that the DNN may suffer from the problem of overfitting which can result in undermined accuracy on unknown test cases.
What is worse, missing synchrophasors introduce noise to the input data, which further compromise the prediction performance.

In order to provide more information for the assessment system, multiple ensemble blocks are employed, each of which comprises two layers of LSTM and a fully-connected hidden neuron layer.
Similarly, the results are post-processed by a $\mathrm{Sigmoid}$ function.
Fig. \ref{fig:Blocks} presents a comparison on the rolled structure of a main block and an ensemble block.

As the ensemble blocks have a much simpler network structure, they are less aware to outliers in the input characteristics at the expense of limited abilities to model complex data with scarce neurons.
Thus only a small portion of all the available synchrophasors are utilized as the inputs of an ensemble block.
Meanwhile, potential overfitting to the chosen PMU data can bias the results.
Multiple ensembles with voltage phasors generated by different PMUs are required to give sufficient secondary results for the decision machine.

In such a system, how to choose voltage phasor collections by PMUs as network input for different ensemble blocks greatly influences the overall prediction accuracy.
To optimally allocate PMU collections for each ensemble block as data input, a PMU input optimization problem is formulated.
The objective function is constructed by considering the system voltage phasor observabilities of multiple sets of PMU.
The overall formulation of the problem is given as follows:
\begin{IEEEeqnarray}{lll} \IEEEyesnumber \label{eqn:ensemble_optim}
&\underset{S_1,\cdots,S_N}{\text{maximize}}\quad \frac{1}{N}(\sum O_i+\sum(O_i-\frac{1}{N}\sum O_i)),\IEEEyessubnumber\\
&\text{subject to}\nonumber\\
&|S_i|=\lfloor P/N\rfloor,\hspace{1em}\forall i\in\{1,2,\cdots,N-1\},\IEEEyessubnumber\label{eqn:4b}\\
&\bigcup S_i=\{1,2,\cdots,P\},\IEEEyessubnumber\label{eqn:4c}\\
&S_i\cap S_j=\varnothing,\hspace{1em}\forall i,j\in\{1,2,\cdots,N\},i\neq j,\IEEEyessubnumber\label{eqn:4d}
\end{IEEEeqnarray}
where $S_i$ is the set of PMUs for ensemble block $i$, $N$ and $P$ are the total numbers of ensemble blocks and PMUs, respectively, and $O_i$ is the voltage phasor observability of PMU set $S_i$.
The constraints in \eqref{eqn:4b} limit the total number of PMUs for each ensemble block.
The constraints in \eqref{eqn:4c} dictate that all PMUs in the system are considered in one ensemble block, and \eqref{eqn:4d} guarantee that each PMU can only be included by one ensemble block.
This combinatorial optimization problem can be solved by using a suitable meta-heuristic and we adopt a recently proposed Social Spider Algorithm (SSA) \cite{yu_social_2015} as the problem solver in this paper.
Note that the problem solver is among the possible techniques to tackle \eqref{eqn:ensemble_optim}.
Alternative methods may be further investigated in future research.

After determining the inputs for each ensemble block, the same training method used in the main block can also be adopted to train these blocks offline.
Meanwhile, online assessments are made only when the input vectors $\mathbf{M}_t$ are available as a whole for ensemble blocks.
The computational expense can be reduced when compared with the main block, and the assessed secondary results are passed to the decision machine for further processing.

\subsection{Decision Machine}\label{sub:rule_machine}

Recall that both the primary and secondary results made by the main and ensemble blocks are values between zero and one.
However, practitioners would prefer a meaningful conclusion on stability instead of statements like ``the system will be 35\% stable''.
Thus it is essential to develop a transformation scheme to map the fuzzy real-valued results into the affirmative 0-1 binary form.

Here we adopt a series of threshold values $\theta_t\in(0,0.5)$ for each block to map the network result $y_t$ to binary values.
Recall that both the main block and the ensemble blocks can generate a sequence of $y_t$ values at different timestamps, each value has a corresponding binary mapping, denoted $z_t$ as follows:
\begin{equation}
z_t = \begin{cases}
1\text{ (Stable)} & \text{for }y_t>1-\theta_t,\\
0\text{ (Unstable)} & \text{for }y_t<\theta_t,\\
?\text{ (Unknown)} & \text{otherwise}.\\
\end{cases}
\end{equation}
The result is considered reliable when we have either $z_t=1$ or $z_t=0$.
The threshold $\theta_t$ of each block in the system is determined offline by solving the threshold optimization problem:
\begin{equation}\label{eqn:threshold_optim}
\underset{\theta_1,\cdots,\theta_T}{\min}\quad(1-A)\times\omega+D-1.
\end{equation}
Here $\omega\in(0,\infty)$ is a weight coefficient, $A$ and $D$ are the testing accuracy and average cycles required to generate a reliable result.
When solving \eqref{eqn:threshold_optim}, the training cases $\{\mathbf{M}_{(c)},y_{(c)}\}$ are employed to generate the value of $A$ using the binary cross entropy error function.
Different values of $\theta$ can result in different 0-1 assessment results $z_t$ from the same $y_t$, thus $A$ is developed from the control variables.
Variable $D$ is calculated with the method introduced in \cite{zhang_post-disturbance_2015}, where proper $\theta$ values help the system make assessments early.
Due to space limitations, the relationship among $\theta$'s, $A$, and $D$ is not provided here.
Interested readers may refer to \cite{zhang_post-disturbance_2015} for more details.
This optimization problem can also be solved by an appropriate metaheuristic (e.g., SSA), and the determined $\theta_t$ values remain constant during the online stability assessment process.

\begin{figure}
\centering
\includegraphics[width=\linewidth]{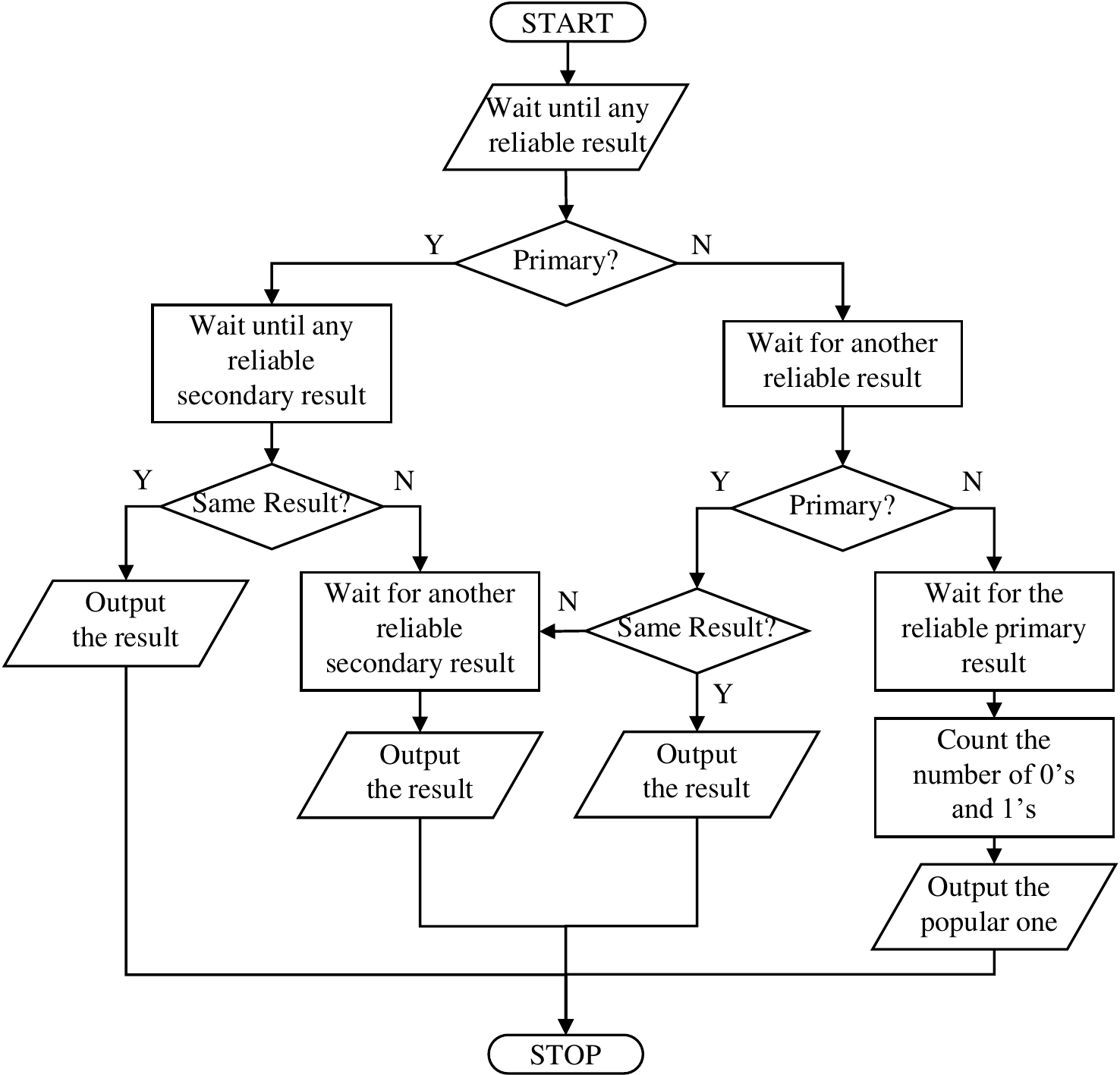}
\caption{Flow chart of generating final assessment result with the rule-based decision machine.}
\label{fig:Decision}
\end{figure}

Besides the above transformation process, another main objective achieved by this decision machine is to combine multiple results and to develop a final assessment result.
The basic idea is to make assessments largely based on the primary result, and use the secondary results for correction.
As the primary and secondary results are generated at different time instants, the decision machine follows the charts given in Fig. \ref{fig:Decision} to consolidate these results.
If the machine receives the primary result, it requires one secondary result for confirmation.
Meanwhile, if one or multiple secondary results are available, the system still need to wait for the arrival of the primary one.
Therefore, we have the following rules:
\begin{itemize}
\item The algorithm starts when the first reliable result is generated by either the main block or any of the ensemble blocks.
\item If the main block makes a reliable assessment first, the algorithm will wait for the first secondary result from any ensemble block.
\item If the secondary result is identical to the primary one, the algorithm outputs the result.
Otherwise, another secondary result is requested, and considered as the final assessment.
\item If the first result is from an ensemble block, the algorithm will wait until the main block generates its assessment.
\item If only one secondary result is made before the primary one, the algorithm compares them and makes a final assessment if they are identical.
Otherwise, another secondary result is requested.
\item If multiple secondary results are made before the primary one, the algorithm uses the more popular result as the final assessment.
\end{itemize}
They are summarized in Fig. \ref{fig:Decision}.
During this whole process, each block can only have one assessment at a time.
New assessments from the same block overwrites the existing one.

\subsection{Work Flow of the Intelligent System}

\begin{figure}
\centering
\includegraphics[width=0.85\linewidth]{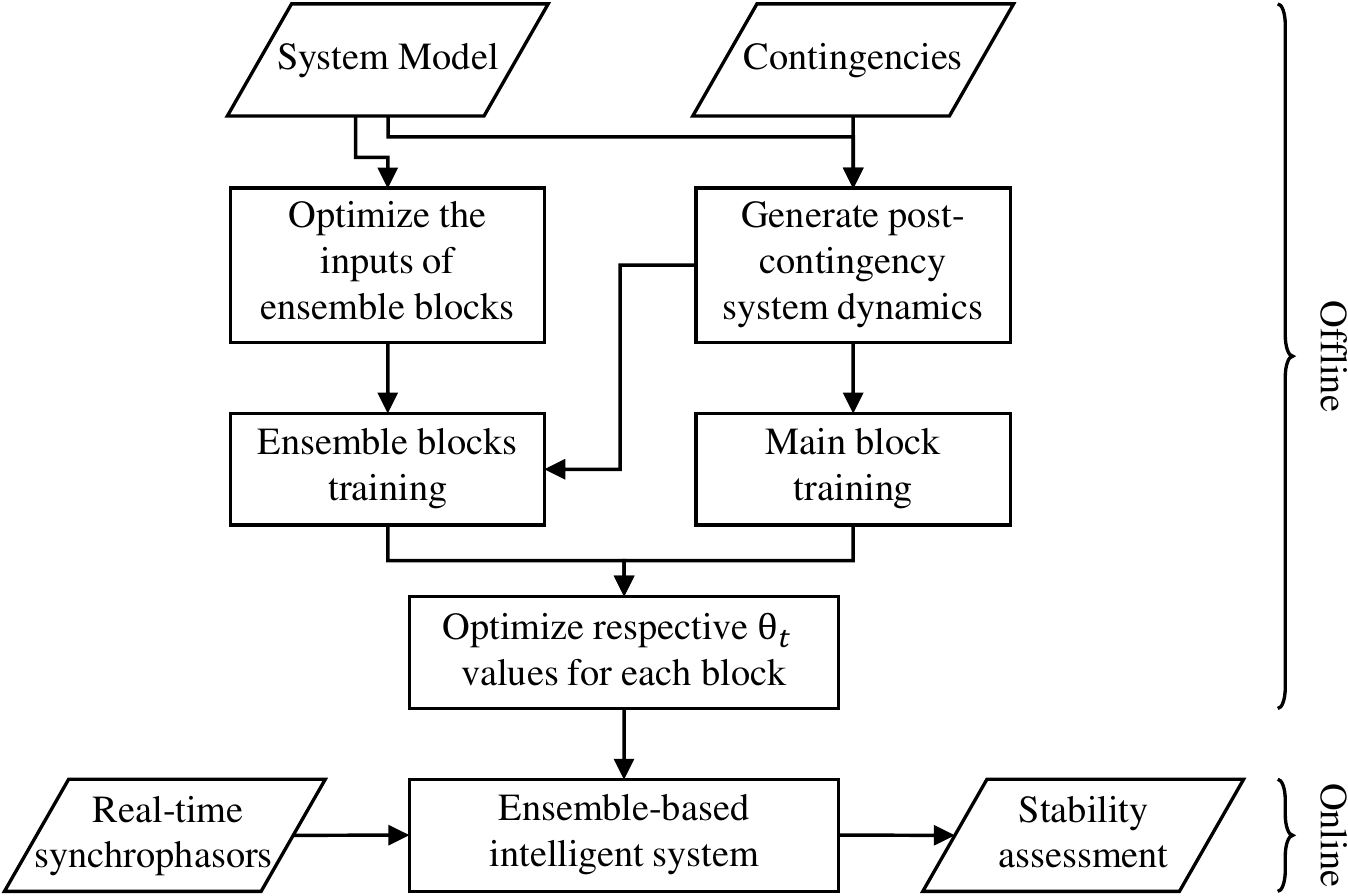}
\caption{Work flow of the proposed intelligent system.}
\label{fig:Infoflow}
\end{figure}

The complete work flow of our proposed system can be divided into two phases as shown in Fig. \ref{fig:Infoflow}, namely, an offline training phase and an online assessment phase.
Utilizing the power system model as well as pre-defined contingency cases, post-contingency system dynamics are calculated in an offline manner.
The dynamics are later employed to train the main block and ensemble blocks of the assessment system.
Meanwhile, the power system topology is considered to develop network inputs of each ensemble block.
After training all blocks in the system, their respective threshold values are optimized using the same input data.
The optimized values as well as the blocks are regarded as the trained ensemble-based intelligent system for the online TSA process.

\begin{figure}
\centering
\includegraphics[width=0.85\linewidth]{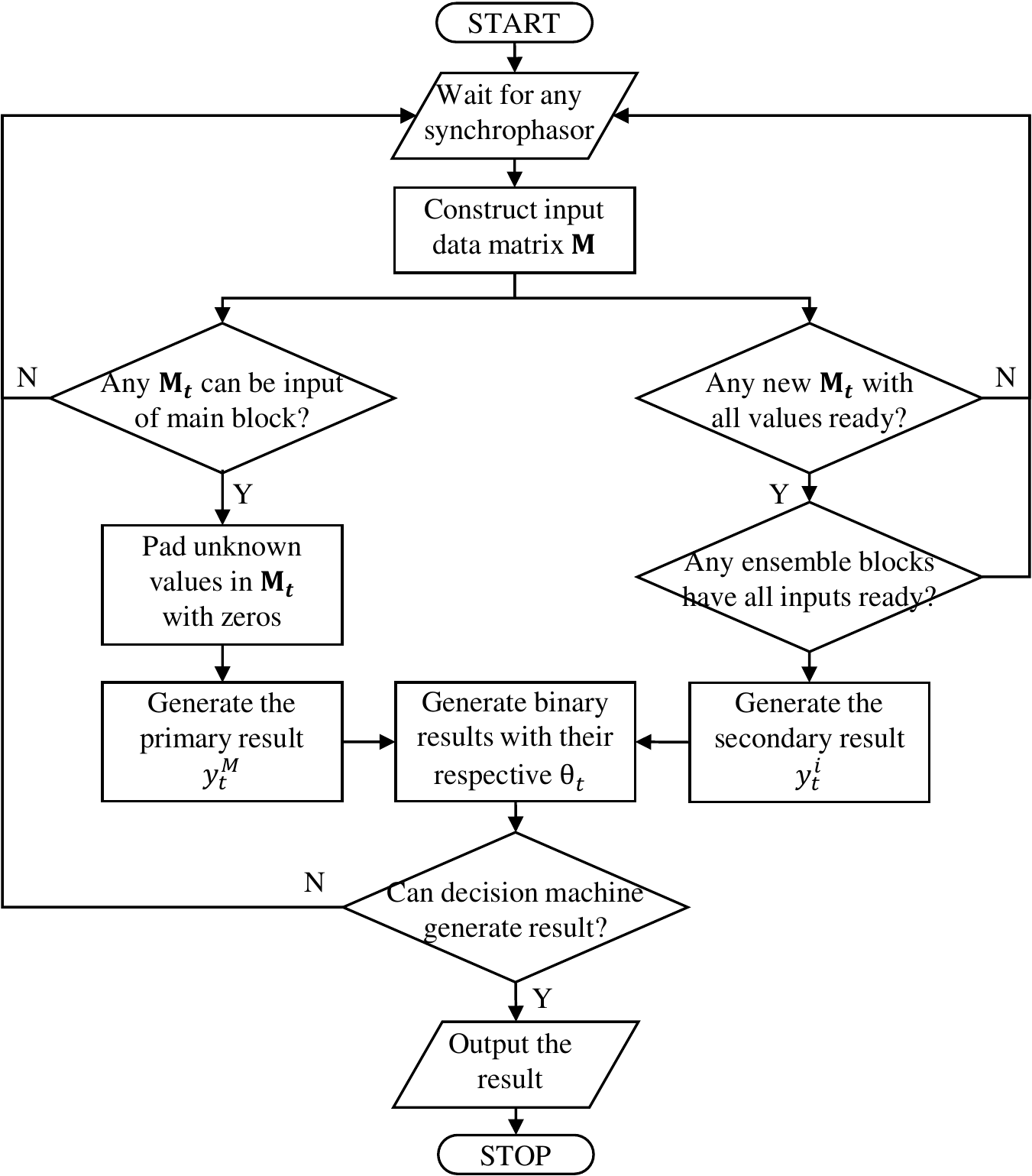}
\caption{Online assessment of the proposed intelligent system.}
\label{fig:Workflow}
\end{figure}

Fig. \ref{fig:Workflow} depicts the online work flow of our proposed ensemble-based intelligent system for TSA.
The assessment is triggered when any new synchrophasor reaches the central controller.
The synchrophasor is placed in the respective position in the input matrix $\mathbf{M}$.
The system then checks if either the main block or any ensemble blocks can generate new results $y_t$ with the incomplete input matrix $\mathbf{M}$.
If so, the generated results are mapped to binary values using pre-optimized $\theta_t$ values.
The calculated $z_t$ results are placed in the decision machine for final assessment generation.
The system stops when the machine outputs an assessment stating either the system will or will not be stable in the future.

\section{Numerical Studies}

We evaluate the performance of the proposed intelligent system through dynamic simulation on the New England 10-machine system benchmark \cite{pai_energy_1989} comprising 39 buses, 10 synchronous generators, 34 transmission lines, and 12 transformers.
It has 19 constant loads consuming a total 6097.1MW and 1408.9 MVAr.
This system is an abstraction of an actual power system in New England \cite{demetriou_dynamic_????}.
Among all generators, G10 represents the aggregated generation from the rest of the eastern interconnection.
All other generators are equipped with an IEEE Type-1 (IEEET1) exciter \cite{_ieee_2006} and a WSCC Type G (BPA\_GG) governor with parameters taken from \cite{demetriou_dynamic_????}.
PMUs are assumed installed on all buses.


\subsection{Training and Testing Preparation}

The collection of training and testing cases are generated by time-domain simulation of post-contingency system dynamics.
These cases comprise the PMU measurements of voltage phasors of all buses in the system and their transient stability 0-1 classifications for selected $N-2$ contingencies.
The loss of any transmission lines or transformers is considered as an $N-1$ contingency.
Then the loss of any remaining component is considered as an $N-2$ contingency.

Furthermore, we consider four operating conditions in which the consumed power is set to 80\%, 90\%, 100\%, and 110\% of the nominal load level, respectively.
A three-phase short-circuit fault is applied at either of the terminal buses of the removed components in $N-2$ contingencies with a fault clearance time of 0.2 seconds.
As a result, 4058 transient contingency cases are created.
All cases are simulated for ten seconds after the fault clearance, and are considered unstable if any generator is out of step.
The time-domain post-contingency system simulation is performed with DIgSILENT PowerFactory \cite{_powerfactory_????}, and all simulations are conducted on an Intel Core i7 CPU at 3.40 GHz clock speed.

The generated transient contingencies are randomly divided into a training set and a testing set.
To comply with the 3:1 training/testing ratio \cite{zhang_post-disturbance_2015}, 3044 cases are used for training the system, while the remaining 1014 are employed to test the system performance.
By adopting this configuration, one can easily tell if the system has over-fit, in which the performance on the training set is superior but that on the testing set is unsatisfactory.

Based on the observation result of \cite{hernandez_one-way_2007}, the transmission delay $\Delta t_{p,t}$ is formulated as a shifted gamma distribution with $k=20$ and $\theta=2.0$.
Although this delay model can only approximate the general one-way delay over the transmission network, real world delay data can be applied to our proposed system and the performance gain shall not be influenced significantly.

\begin{table}
	\caption{Simulation Parameters}
    \centering
	\begin{tabular}{ll|ll|ll}
	\hline
    \multicolumn{6}{c}{System Parameters} \\
    \hline
    $\phi$ & $0.5$ & $N$ & $3$ to $7$ & $\omega$ & $100$ \\
    \hline
    \hline
    \multicolumn{6}{c}{SSA Parameters} \\
    \hline
    $|pop|$ & $30$ & $max\_iter$ & $2000$ & $r_a$ & $1.0$ \\
    $p_c$ & $0.7$ & $p_m$ & $0.1$ & & \\
    \hline
	\end{tabular}
	\label{tbl:params}
\end{table}

All system parameters are listed in Table \ref{tbl:params}.
Among these defined parameters, the number of ensemble blocks $N$ are set to five values, i.e., $3$ to $7$.
This means that in the test, Problem \eqref{eqn:ensemble_optim} is optimized five times and each optimization generates $N$ blocks.
As a result, $3+4+5+6+7=25$ ensemble blocks are created to facilitate the generation of secondary results.

\subsection{Impact of Delayed Measurements} \label{sub:comp}

We first compare the proposed delay aware TSA mechanism, labeled by ``Delay aware TSA'' in the sequel, with conventional techniques in which the assessments are generated after receiving all measurements.
For fair comparison, we employ the existing fastest TSA mechanism proposed in \cite{zhang_post-disturbance_2015}, labeled by ``Synchronous TSA (STSA)'', for performance assessment.
The communication latency values for both mechanisms are identical, and STSA is carried out when the control center can calculate the complete system state.
In addition, we further assume that STSA can generate assessments with information of the first post-contingency cycle, and the calculation is instantaneous.

\begin{table}
  \caption{Comparison of Assessment Accuracy and Response Time}
	\label{tbl:is_stsa}
	\begin{tabular}{r|ccc|cc}
		\hline
		\multirow{2}*{Mechanism} & \multicolumn{3}{c|}{Response Time (ms)} & \multicolumn{2}{c}{Accuracy (\%)} \\
		\cline{2-6}
		& Average & Best & Worst & Training & Testing \\
		\hline
		Delay aware TSA & \textbf{48.0} & \textbf{35.4} & \textbf{79.9} & \textbf{99.8} & \textbf{100.0} \\
		Synchronous TSA & $>$82.6 & $>$58.9 & $>$141.3 & N/A & 99.4 \\
		\hline
	\end{tabular}
	\centering
\end{table}

The simulation results are presented in Table \ref{tbl:is_stsa}, where the better performing results are in bold.
In this table, the assessment accuracy and response times for both mechanisms are presented.
The response times for synchronous TSA in practice are always greater than the listed value, thus prepended with ``$>$'' signs.
It can be concluded that delayed measurements have a significant impact on the TSA response time.
On average, the proposed delay aware TSA can achieve around $1.7$x speedup than state-of-the-art conventional TSA mechanism.
In addition, thanks to all the block networks in delay aware TSA, the proposed mechanism generates correct assessments in all test cases, and can provide almost perfect accuracy in the training cases.

For completeness of performance assessment, we also measure the average training time for the proposed delay aware TSA.
On average, the training time for the main block is 1341 seconds, and that for each of the ensemble blocks is 176.2 seconds.
Therefore, the whole system can be trained within 5746 seconds sequentially.
Moreover, as the training of different blocks are independently, it is simple and effective to train the networks paralleled.
In such a case, the proposed system can adapt to significant changes in operating conditions in little time.
Note that the training data already contains different operating conditions.
So insignificant changes can be addressed using the same network without re-training.

\subsection{Assessment Accuracy and Response Time}

As presented in Section \ref{sub:comp}, the proposed delay aware TSA can achieve a superior performance compared with conventional mechanisms.
It is of interest to determine which component(s) of the mechanism contribute to the performance.
In this test, we separate the delay aware TSA mechanisms into two sub-systems, composed of the main block and all ensemble blocks, respectively.
For the ensemble blocks sub-system, final assessment is made when at least two ensembles generate identical secondary assessment results.
In addition, we also investigate the impact of measurement inclusion parameter $\phi$.

\begin{table}
  \caption{Comparison of Different Variants of Proposed Delay Aware TSA Mechanism}
	\label{tbl:block_performance}
	\begin{tabular}{r|ccc|cc}
		\hline
		\multirow{2}*{Mechanism} & \multicolumn{3}{c|}{Response Time (ms)} & \multicolumn{2}{c}{Accuracy (\%)} \\
		\cline{2-6}
		& Average & Best & Worst & Training & Testing \\
		\hline
		Delay aware TSA & \textbf{48.0} & 35.4 & 79.9 & \textbf{99.8} & \textbf{100.0} \\
		Main Block, $\phi=0.5$ & 48.9 & \textbf{31.2} & 93.7 & 94.7 & 92.9 \\
		Ensemble Blocks & 49.5 & 36.6 & \textbf{79.3} & 99.5 & 98.4 \\
		\hline
		Main Block, $\phi=1.0$ & 85.4 & 46.3 & 144.7 & \textbf{99.7} & \textbf{99.4} \\
		Main Block, $\phi=0.8$ & 70.9 & 41.5 & 104.3 & 98.0 & 97.5 \\
		Main Block, $\phi=0.6$ & 56.1 & 35.0 & 89.3 & 95.1 & 93.8 \\
		Main Block, $\phi=0.4$ & \textbf{45.7} & \textbf{27.9} & \textbf{65.7} & 91.9 & 92.4 \\
		\hline
	\end{tabular}
	\centering
\end{table}

The simulation results are compared in Table \ref{tbl:block_performance}.
It can be seen that neither main block nor ensemble blocks can generate perfect assessment results, but combining them together with the decision machine yields superior performance.
In addition, the proposed mechanism can have a slightly shorter average response time than either of the sub-systems.
So we can conclude that the system performance is contributed to by all major components of the delay aware TSA, namely the main block, ensemble blocks, and the decision machine.

In addition, it can be summarized that the parameter $\phi$ plays an important role in balancing the trade-off between assessment accuracy and response time.
While a larger $\phi$ leads to more accurate assessments, a small $\phi$ makes the system wait for less measurements, rendering a faster response speed.

\begin{figure}
\centering
\includegraphics[width=\linewidth]{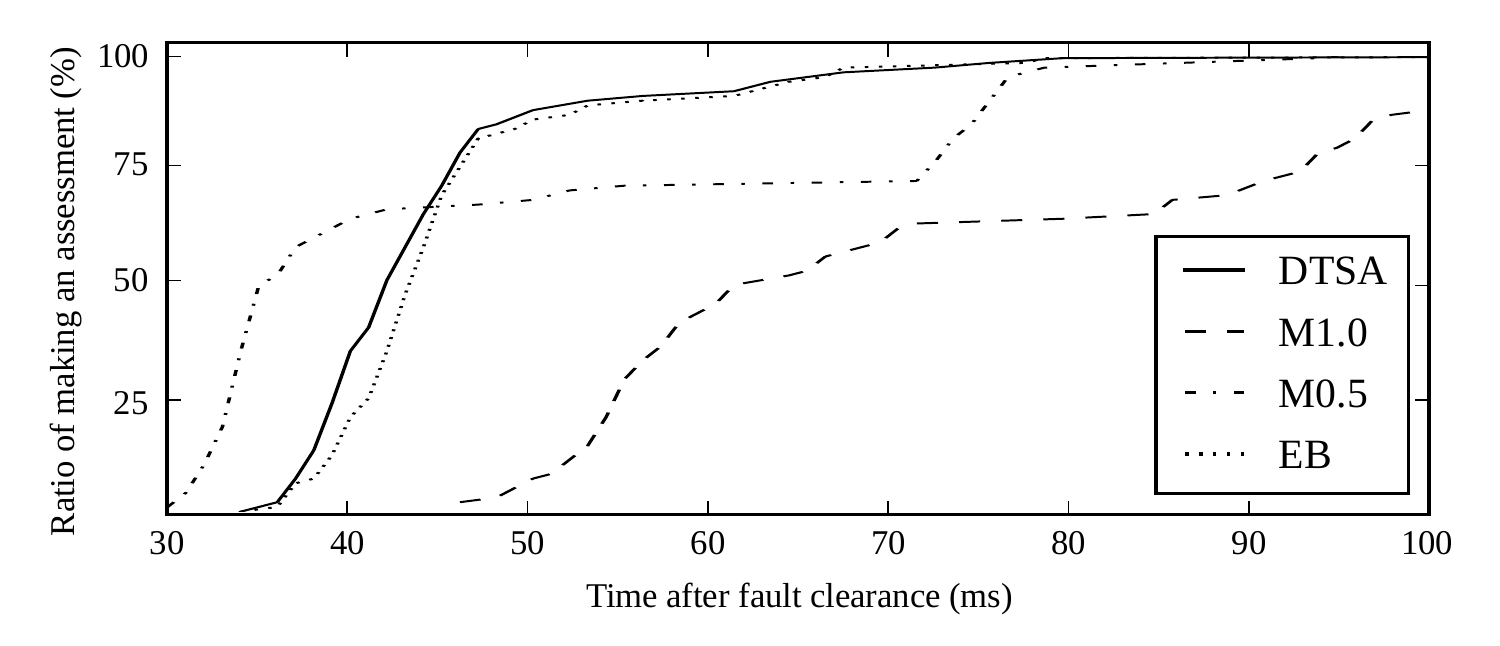}
\caption{Rate of assessments with respect to response time.
``DTSA'' is the proposed delay aware TSA mechanism, ``M1.0'' is the main block sub-system with $\phi=1.0$, ``M0.5'' is the main block sub-system with $\phi=0.5$, and ``EB'' is the ensemble blocks sub-system.}
\label{fig:Comparison}
\end{figure}

The rates of assessments with respect to response time are depicted in Fig. \ref{fig:Comparison} for a better understanding of the results.
The figure depicts that the proposed delay aware TSA can generally develop most assessment in the shortest average response time.
It can also be observed that while main block sub-system with $\phi=0.5$ can start making assessment earliest among all mechanisms, it suffers from random latency spikes in some test cases and has to wait for the delay measurements in these cases.
Both conclusions accord with the results in Table \ref{tbl:block_performance}, and demonstrate the superiority of the proposed mechanism.


\subsection{Transient Assessment on Noisy-Delayed Measurements}

All previous simulation assumes that the PMU measurements, whenever they arrive at the central controller, can represent the system dynamic state accurately.
However, in practice these measurements may be noisy.
Therefore, additional numerical simulations are carried out to study the influence of noisy and delayed measurements on the proposed system for TSA.

According to IEEE Standard for Synchrophasor Data Transfer for Power Systems (C37.118.2-2011) \cite{_ieee_2011}, all PMUs complying with the standard shall generate system variable measurements with a total vector error less than 1\%.
Thus in this paper we follow the approach introduced in \cite{he_online_2013} to generate noisy test cases:
\begin{equation}
\tilde{V}\tilde{\angle\theta_V} = V\angle\theta_V + \Delta V\angle\Delta\theta_V,
\end{equation}
where $\tilde{V}\tilde{\angle\theta_V}$ is the measured voltage phasor, $V\angle\theta_V$ is the actual voltage phasor, and $\Delta V\angle\Delta\theta_V$ is the noise phasor imposed, which satisfies a truncated complex normal distribution \cite{he_online_2013}.
The newly generated test cases are employed to test the assessment performance of the proposed system trained using noiseless training cases.

\begin{table}
  \caption{Comparison of Noisy and Noiseless Data}
	\label{tbl:noisy}
	\begin{tabular}{r|ccc|cc}
		\hline
		\multirow{2}*{Mechanism} & \multicolumn{3}{c|}{Response Time (ms)} & \multicolumn{2}{c}{Accuracy (\%)} \\
		\cline{2-6}
		& Average & Best & Worst & Training & Testing \\
		\hline
		Delay aware TSA & \textbf{48.0} & 35.4 & \textbf{79.9} & \textbf{99.8} & 100.0 \\
		Noisy Data & 48.2 & \textbf{31.7} & 83.5 & \textbf{99.8} & \textbf{99.9} \\
		\hline
	\end{tabular}
	\centering
\end{table}

\begin{figure}
\centering
\includegraphics[width=\linewidth]{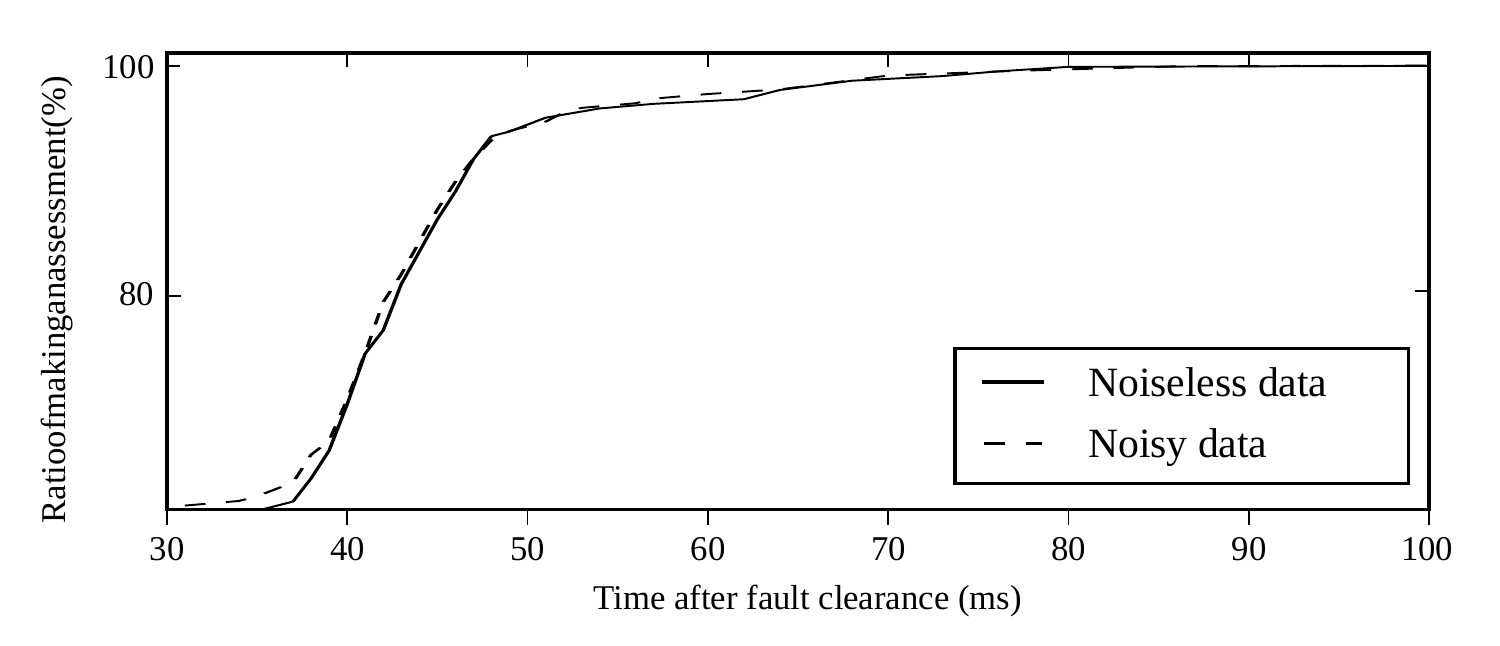}
\caption{Performance of proposed intelligent system on making early assessments with noiseless and noisy data.}
\label{fig:Noisy}
\end{figure}

The assessment result is summarized in Table \ref{tbl:noisy} and Fig. \ref{fig:Noisy}.
It can be observed from the comparison that data noise makes trivial influence on the assessment accuracy and response time.
This is contributed by the outstanding classification ability of neural networks on noisy data \cite{lippmann_introduction_1987,kingma_adam:_2015}, and the introduction of multiple networks to form an ensemble \cite{hansen_neural_1990}.

\section{Conclusion}

In this paper, an intelligent system is proposed to address the data transmission delay in an online TSA process.
This system is based on Long-short Term Memory (LSTM) ensemble neural network with strategically designed decision machine.
In particular, one large LSTM network called main block and multiple small LSTM networks called ensemble blocks cooperate to provide a collection of transient assessments considering different input data for the decision machine to make a final TSA conclusion.
Different from existing TSA work, the proposed system can adapt to the delayed PMU measurements and make reliable assessments at the earliest possible time to facilitate later control actions.
Simulation results show that the developed system outperforms conventional TSA methodologies in terms of the average response time and maintains a perfect assessment accuracy.
In addition, the simulation demonstrates that both the main block and ensemble blocks contribute to superior TSA performance.
The proposed system is also tested with noisy system measurement data, and the result indicates that the system is robust with noisy data complying with the related IEEE standard.

The contributions of this paper are summarized as follows:
\begin{itemize}
\item We propose a delay aware transient stability assessment to address the impact of communication delay in the process of transient stability assessment.
The delay is critical for fast response time, but has not been considered in the previous literature.
\item We develop an LSTM ensemble-based intelligent system to handle the assessment problem with delayed and missing data caused by communication delay.
The system can be further extended to more power system data-driven applications.
\item We assess the system on a widely adopted testbed, and provide configuration guidelines to fully utilize its assessment capability.
The simulation results demonstrate a significant improvement in system response time while maintaining a perfect accuracy.
\end{itemize}
Future efforts will focus on the availability of PMUs in the system.
It is assumed that PMUs are installed on all available buses in the system, which may be practical due to the decreasing PMU prices.
However it is still of interest to find the minimal number and locations of PMUs needed to fulfill the assessment task.
Besides, different assessment predictors may contribute to a better average response time.

\bibliography{zotero}
\bibliographystyle{IEEEtran}

\begin{IEEEbiography}[{\includegraphics[width=1in,height=1.25in,clip,keepaspectratio]{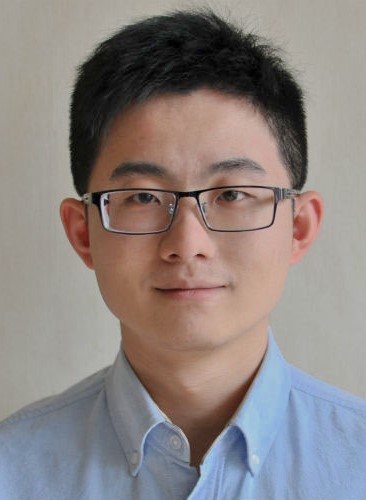}}]{James J.Q. Yu}(S'11--M'15)
received the B.Eng. and Ph.D. degree in Electrical and Electronic Engineering from the University of Hong Kong, Pokfulam, Hong Kong, in 2011 and 2015, respectively. He is currently a postdoctoral fellow at the Department of Electrical and Electronic Engineering of the University of Hong Kong. His research interests include smart city technologies, power stability analysis, evolutionary algorithm design and analysis, and deep neural network applications.
\end{IEEEbiography}

\begin{IEEEbiography}[{\includegraphics[width=1in,height=1.25in,clip,keepaspectratio]{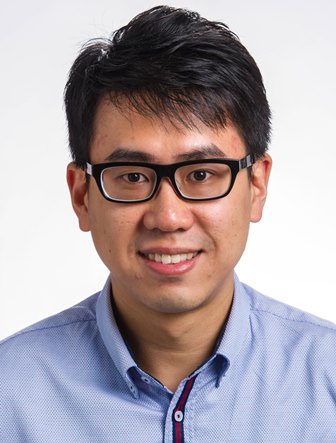}}]{Albert Y.S. Lam}(S'03--M'10--SM'16)
received the BEng degree (First Class Honors) in Information Engineering and the PhD degree in Electrical and Electronic Engineering from the University of Hong Kong (HKU), Hong Kong, in 2005 and 2010, respectively. He was a postdoctoral scholar at the Department of Electrical Engineering and Computer Sciences of University of California, Berkeley, CA, USA, in 2010--12, and now he is a research assistant professor at the Department of Electrical and Electronic Engineering of HKU. He is a Croucher research fellow. His research interests include optimization theory and algorithms, evolutionary computation, smart grid, and smart city.
\end{IEEEbiography}

\begin{IEEEbiography}[{\includegraphics[width=1in,height=1.25in,clip,keepaspectratio]{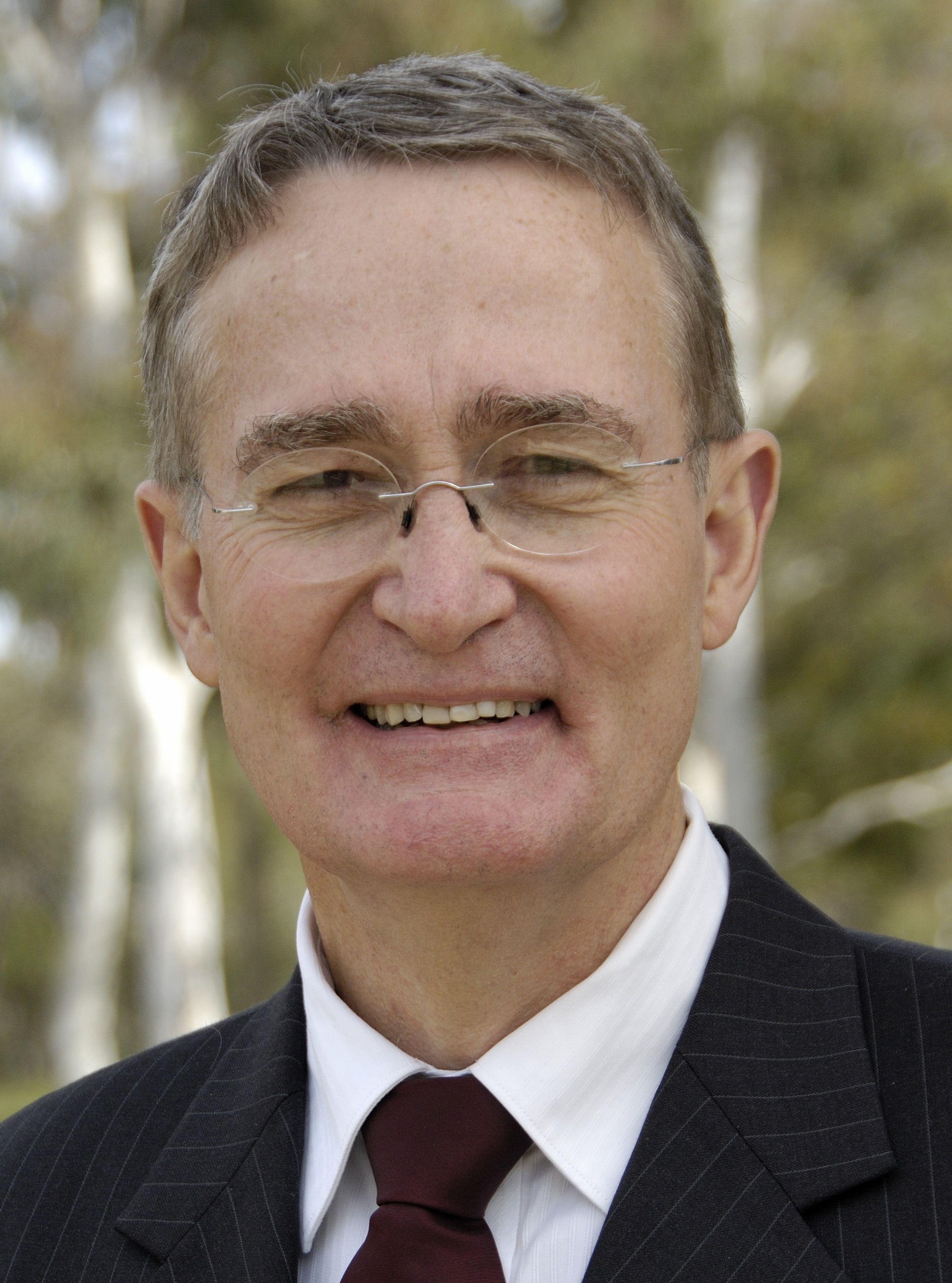}}]{David J. Hill}(S'72-M'76-SM'91-F'93-LF'14) received the PhD degree in Electrical Engineering from the University of Newcastle, Australia, in 1976. He holds the Chair of Electrical Engineering in the Department of Electrical and Electronic Engineering at the University of Hong Kong. He is also a part-time Professor of Electrical Engineering at The University of Sydney, Australia. 
During 2005-2010, he was an Australian Research Council Federation Fellow at the Australian National University. Since 1994, he has held various positions at the University of Sydney, Australia, including the Chair of Electrical Engineering until 2002 and again during 2010-2013 along with an ARC Professorial Fellowship. He has also held academic and substantial visiting positions at the universities of Melbourne, California (Berkeley), Newcastle (Australia), Lund (Sweden), Munich and in Hong Kong (City and Polytechnic). 

His general research interests are in control systems, complex networks, power systems and stability analysis. His work is now mainly on control and planning of future energy networks and basic stability and control questions for dynamic networks.
Professor Hill is a Fellow of the Society for Industrial and Applied Mathematics, USA, the Australian Academy of Science, the Australian Academy of Technological Sciences and Engineering and the Hong Kong Academy of Engineering Sciences.. He is also a Foreign Member of the Royal Swedish Academy of Engineering Sciences.
\end{IEEEbiography}

\begin{IEEEbiography}[{\includegraphics[width=1in,height=1.25in,clip,keepaspectratio]{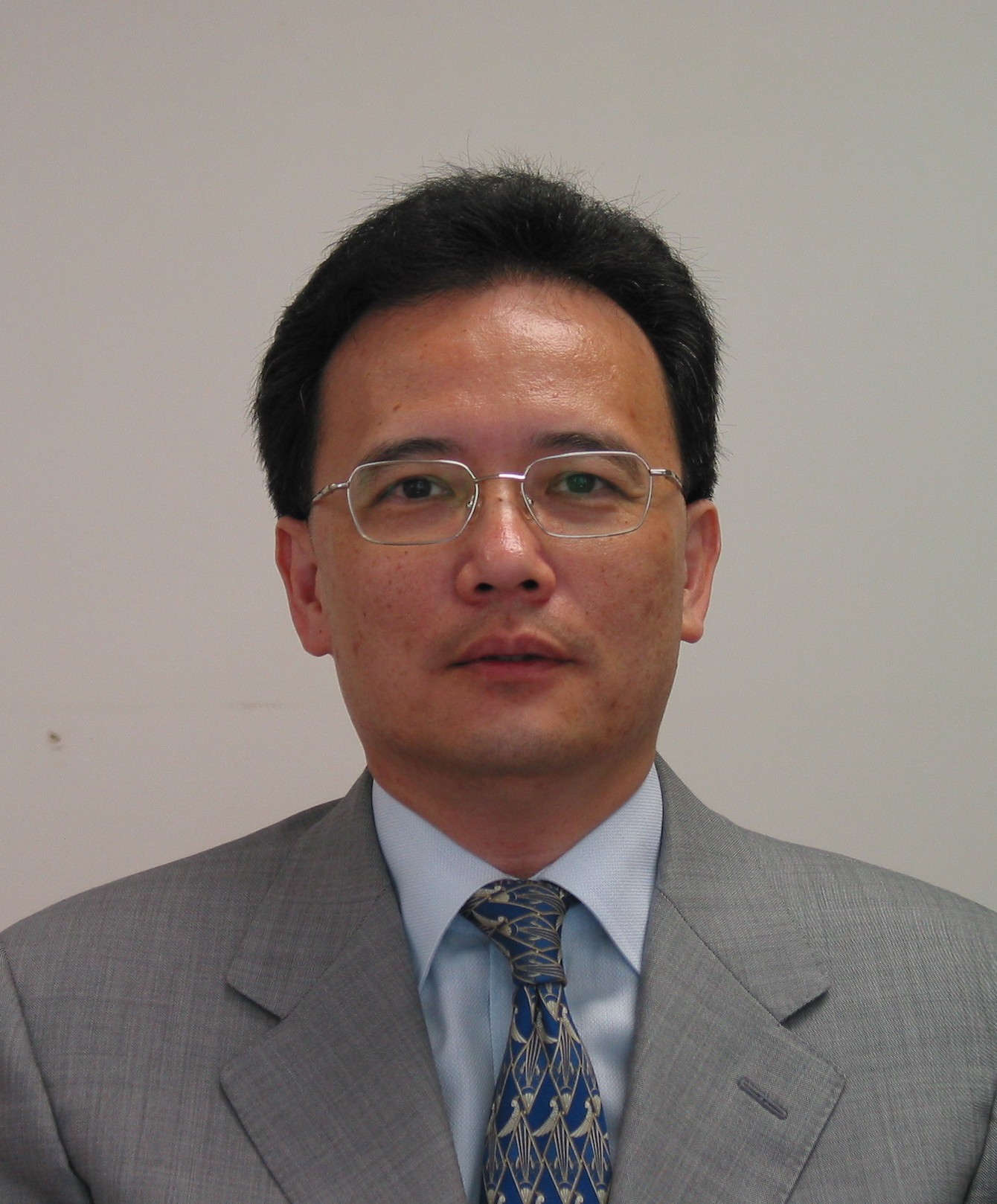}}]{Victor O.K. Li}(S'80-M'81-SM'86-F'92) received SB, SM, EE and ScD degrees in Electrical Engineering and Computer Science from MIT in 1977, 1979, 1980, and 1981, respectively. He is Chair of Information Engineering, Cheng Yu-Tung Professor in Sustainable Development, and Head of the Department of Electrical and Electronic Engineering at the University of Hong Kong (HKU). He has also served as Assoc. Dean of Engineering and Managing Director of Versitech Ltd., the technology transfer and commercial arm of HKU. He served on the board of China.com Ltd., and now serves on the board of Sunevision Holdings Ltd., listed on the Hong Kong Stock Exchange. Previously, he was Professor of Electrical Engineering at the University of Southern California (USC), Los Angeles, California, USA, and Director of the USC Communication Sciences Institute. His research is in the technologies and applications of information technology, including clean energy and environment, social networks, wireless networks, and optimization techniques. Sought by government, industry, and academic organizations, he has lectured and consulted extensively around the world. He has received numerous awards, including the PRC Ministry of Education Changjiang Chair Professorship at Tsinghua University, the UK Royal Academy of Engineering Senior Visiting Fellowship in Communications, the Croucher Foundation Senior Research Fellowship, and the Order of the Bronze Bauhinia Star, Government of the Hong Kong Special Administrative Region, China.  He is a Registered Professional Engineer and a Fellow of the Hong Kong Academy of Engineering Sciences, the IEEE, the IAE, and the HKIE.
\end{IEEEbiography}

\end{document}